\documentclass[prb,aps,amsmath,amssymb,amsfonts,floatfix,superscriptaddress,showpacs,twocolumn,footinbib]{revtex4-1}
\usepackage{graphicx}
\usepackage[dvipsnames]{xcolor}
\usepackage{bbm}
\usepackage[utf8]{inputenc}
\usepackage{dcolumn}
\usepackage{hyperref}
\usepackage[caption=false]{subfig}
\usepackage[english]{babel}
\usepackage{indentfirst}
\usepackage{textcomp}
\usepackage{comment}
\hypersetup{colorlinks=true,breaklinks,linkcolor=blue,urlcolor=blue,citecolor=blue}

\usepackage{tikz}
\usetikzlibrary{calc}
\usetikzlibrary{decorations.pathmorphing}
\usetikzlibrary{decorations.pathreplacing}
\usetikzlibrary{decorations.markings}

\tikzset{->-/.style={decoration={
  markings,
  mark=at position #1 with {\arrow{>}}},postaction={decorate}}}
\tikzset{-<-/.style={decoration={
  markings,
  mark=at position #1 with {\arrow{<}}},postaction={decorate}}}

\usepackage{lipsum, babel}

\newcommand{\kv}{\ensuremath{\mathbf{k}}}

\newcommand{\qv}{\ensuremath{\mathbf{q}}}

\newcommand{\dv}{\ensuremath{\hat{d}}} 

\newcommand{\av}[1]{\ensuremath{\left\langle #1 \right\rangle}}

\def \half {\ensuremath{\frac{1}{2}}}

\begin{document}

\title{Interaction-driven Lifshitz transition with dipolar fermions in optical lattices}

\author{E. G. C. P. van Loon}
\affiliation{Radboud University, Institute for Molecules and Materials, NL-6525 AJ Nijmegen, The Netherlands}

\author{M. I. Katsnelson}
\affiliation{Radboud University, Institute for Molecules and Materials, NL-6525 AJ Nijmegen, The Netherlands}

\author{L. Chomaz}
\affiliation{Institut f\"{u}r Experimentalphysik, Universit\"{a}t Innsbruck, Technikerstraße 25, 6020 Innsbruck, Austria}
\affiliation{Institut f\"{u}r Quantenoptik und Quanteninformation, \"{O}sterreichisches Akademie der Wissenschaften, 6020 Innsbruck, Austria}

\author{M. Lemeshko}
\email{mikhail.lemeshko@ist.ac.at}
\affiliation{IST Austria (Institute of Science and Technology Austria), Am Campus 1, 3400 Klosterneuburg, Austria} 

\begin{abstract} 
Anisotropic dipole-dipole interactions between ultracold dipolar fermions break the symmetry of the Fermi surface and thereby deform it. Here we demonstrate that such a Fermi surface deformation induces a topological phase transition -- so-called Lifshitz transition -- in the regime accessible to present-day experiments. We describe the impact of the Lifshitz transition on observable quantities such as the Fermi surface topology, the density-density correlation function, and the excitation spectrum of the system. 
The Lifshitz transition in ultracold atoms can be controlled by tuning the dipole orientation and -- in contrast to the transition studied in crystalline solids -- is completely interaction-driven.
\end{abstract}


\maketitle

\section{Introduction}

The concept of Fermi surface plays a central role in the description of electronic systems. Several physical properties, such as the electrical conductivity and the absorption spectrum of the system are determined by the shape of the Fermi surface~\cite{Lifshitz56,Abrikosov88,Mahan00} as well as the electrons' dispersion relation in the vicinity of this surface. 
Stationary points of the dispersion relation correspond to the Van Hove singularities (VHS)~\cite{VanHove53}. If a VHS occurs close to the Fermi surface, it can dramatically alter the properties of the electron gas. 
For example, in two-dimensional systems the density of states exhibits a logarithmic divergence at the VHS~\cite{VanHove53}.
If one deforms the Fermi surface such that it crosses a VHS, there occurs an electronic topological transition -- the Lifshitz transition~\cite{Lifshitz60,Lifshitz56,Blanter1994,Katsnelson94}.

The Lifshitz transition has been explored in a variety of systems, from high-temperature copper-oxide~\cite{Irkhin01,Irkhin02,Botelho05,Norman10} and iron-based superconductors~\cite{Leonov15} to superfluid helium~\cite{Silaev15}.  In condensed matter systems, the change in Fermi surface at the Lifshitz transition affects observable quantities such as resistivity and thermoelectric power~\cite{Blanter1994,Vaks81}, lattice dynamics, elastic
moduli and related thermal properties such as heat capacity and thermal expansion~\cite{Katsnelson94,Varyukhin1988,Potzel95,Kenichi95,Novikov99,Vaks89,Vaks91,Souvatzis07,Glazyrin13}.
In some cases, it determines the peculiarities of phase diagrams of metals under pressure as well as metal alloys~\cite{Katsnelson94,Vaks89,Vaks91}.
In such settings, there is a strong and complicated interplay between the electrons experiencing the transition and the underlying ionic lattice. In isotropic systems, it is challenging to induce the Lifshitz transition using a tunable interaction, since Luttinger's theorem~\cite{Luttinger1960,Luttinger1960-2} combined with the symmetry of the system strongly constrains the Fermi surface.\footnote{A spontaneous symmetry breaking (Pomeranchuk instability) scenario is possible~\cite{Bunemann12}. Contrary to the Pomeranchuk scenario, the Lifshitz transition with dipolar fermions as discussed here is continuously tunable since the symmetry is broken explicitly by the dipolar interaction.}  Therefore, usually the Lifshitz physics is studied by changing the single-particle properties of the system, such as the chemical potential or the electrons' kinetic energy~\cite{Chen12,Slizovskiy14}. 

Experiments with ultracold atomic and molecular Fermi gases in optical lattices pave the way to unravel the properties of strongly-correlated condensed-matter systems using ``clean'' and highly tunable setups~\cite{Lewenstein2007,RevModPhys.80.885, RevModPhys.86.153}, exemplifying the concept of a quantum simulator as introduced by Feynman\cite{Feynman82}. For instance, it became possible to prepare a fermionic Mott insulator~\cite{Schneider2008,Jordens2008} and study the properties of the repulsive Fermi-Hubbard model~\cite{Esslinger2010, Kaczmarczyk16}, probe the BEC-BCS crossover in lattices \cite{Chin2006}, study short-range magnetism~\cite{GreifScience13} and multiflavor spin dynamics \cite{KrauserNatPhys12}, as well as to realize artificial graphene sheets~\cite{UehlingerPRL13} and the topological Haldane model~\cite{JotzuNature14}. 

Typical ultracold fermion experiments deal with short-range isotropic interparticle interactions. Recent experimental efforts, however, have been devoted to exploit particles possessing a large electric or magnetic dipole moment.
Ultracold fermionic molecules, such as $^{40}$K$^{87}$Rb~\cite{OspelkausFDiss09, NiNature10} and $^{23}$Na$^{40}$K~\cite{ParkPRL15, Park15b}, have been prepared in their absolute ground states, while ultracold gases of magnetic atoms such as $^{161}$Dy~\cite{LuPRL12}, $^{167}$Er~\cite{AikawaPRL14}, and $^{53}$Cr~\cite{NaylorPRA15}, have been brought to Fermi degeneracy. As opposed to the conventional condensed matter systems where the Coulomb interaction between electrons is screened by the ionic crystal, these systems allow to realize truly long-range interactions between the trapped  particles.
One further advantage of ultracold gases compared to condensed matter systems is their high tunability. For instance, the relative strength of the long- and short-range interactions can be controlled via Feshbach resonances\cite{Inouye:1998,Courteille:1998} and control over the long-range interaction via time-dependent dipole orientation~\cite{Giovanazzi:2002} or state-dressing~\cite{Buechler:2007}.
The anisotropic and long-range character of the dipole-dipole interactions (DDI) is predicted to give rise to novel many-body Hamiltonians~\cite{BaranovCRev12, KreStwFrieColdMolecules, CarrNJP09, LemKreDoyKais13, LahayePfauRPP2009, BaranovPRep08, TrefzgerJPB11}, some of which have already been realized in laboratory~\cite{dePazPRL13, YanNature13, Baier201}. 

In this work we demonstrate that dipolar quantum gases trapped in optical lattices offer a unique opportunity to study the physics associated with Lifshitz transitions. First, the ultracold experimental setups allow to tune the properties of the fermions and  underlying lattice independently, which is rather challenging to realize in crystalline solids. 
Second, the anisotropic nature of DDI breaks the spatial symmetry of the system, manifesting itself in Fermi surface deformations~\cite{Miyakawa08,Fregoso09,Baillie12},  as recently observed in experiment~\cite{Aikawa14}. Here we show that, in the context of lattice systems, such deformations can be used to generate a   Lifshitz  transition, which, in turn,  has a strong impact on the correlations in the system. Since the orientation of the dipoles can be controlled by an external field, dipolar fermions provide a convenient way to study such a transition experimentally. Juxtaposed to the Lifshitz transition observed in solids, the one studied here is \textit{interaction-driven}, i.e.\ it occurs solely due to the two-particle terms of the Hamiltonian.

A similar scenario has been investigated theoretically in coupled quasi-1D chains of ultracold atoms~\cite{Quintanilla09,Carr10}. There, the interchain hopping was used as the tuning parameter and the external field was oriented to rule out intrachain interactions.  In contrast, the transition studied in this paper occurs in an isotropic lattice, and the dipolar character of the fermions is truly essential.

\section{Dipolar fermions on an optical lattice}

\begin{figure}
 \includegraphics[width=0.48\textwidth]{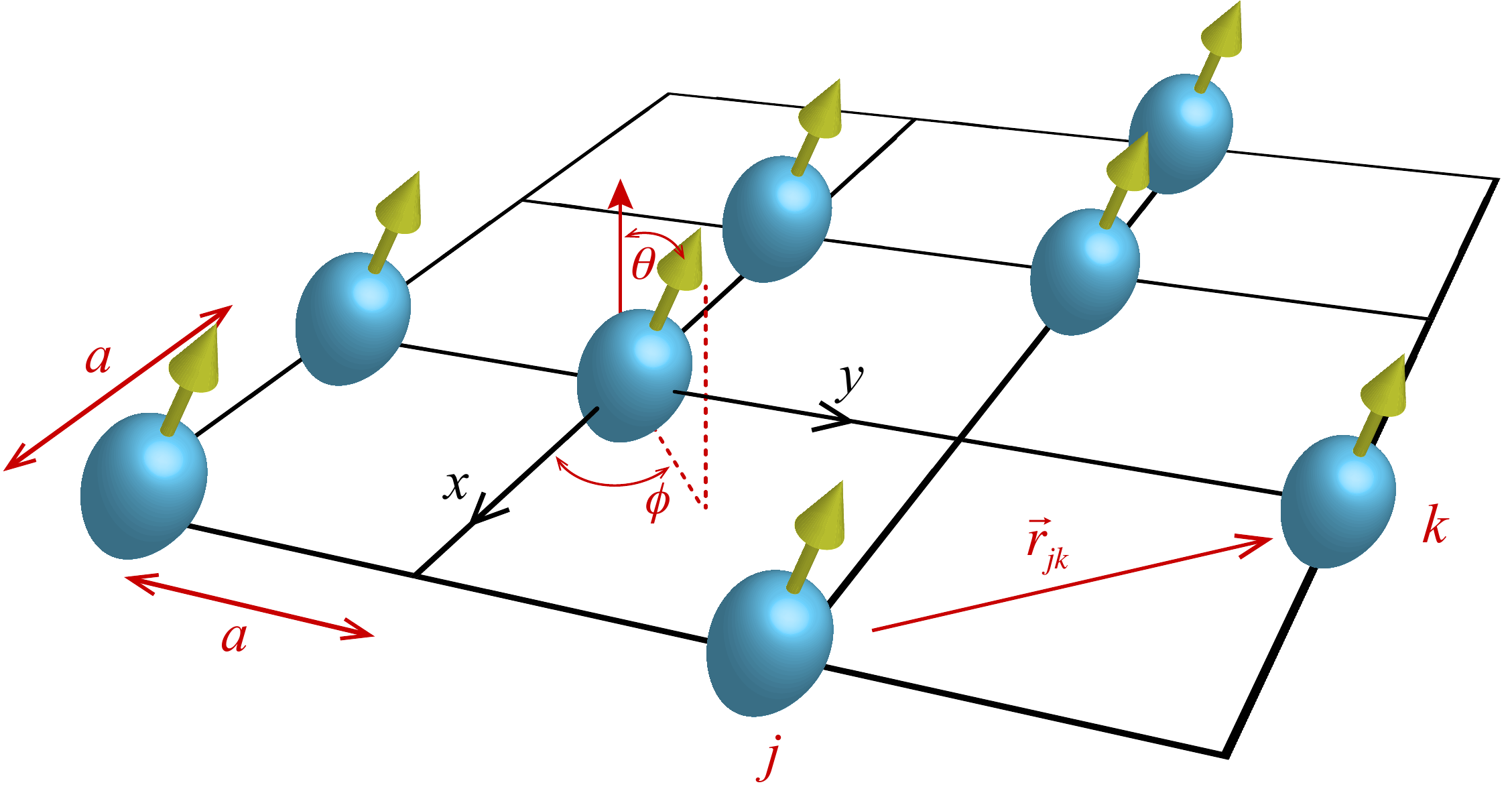}
 \caption{Single-component dipolar fermions on a square optical lattice. Due to the symmetry, the orientation of the dipoles is given by two angles, $\phi \in [0,0.25\pi]$ and $\theta \in [0,0.50\pi]$.
 }
 \label{fig:lattice}
\end{figure}

We expect the physics of the Lifshitz transition to be qualitatively similar for any Hubbard-like model in the Fermi liquid phase. 
Therefore, without loss of generality, we restrict ourselves to the single-component dipolar fermion model on a square two-dimensional lattice, as schematically illustrated in Fig.~\ref{fig:lattice}. Furthermore, ultracold atomic gases of fully polarized fermions are readily available in experiment~\cite{LuPRL12, AikawaPRL14, NaylorPRA15}, and allow to avoid dealing with complex spin preparation protocols and dipolar relaxation effects between spin components that modifies the initial spin preparation\cite{Hensler2003,Fattori2006dc,Pasquiou2010cod,Pasquiou2011sra,Pasquiou2011sdo, dePaz2013rdd,Lu2010, Newman:2011,Burdick:2015}. 

The model's Hamiltonian is given by:
\begin{align}
\label{eq:hmlt}
  H = - t \sum_{\av{jk}} c_{j}^\dagger c_{k}^{\phantom{\dagger}} + \half \sum_{jk} V^d_{jk} n_j n_k,
\end{align}
where $c_{j}^\dagger$ ($c_{j}^{\phantom{\dagger}}$) creates (annihilates) a fermion on site $j$, and $n_j = c_{j}^\dagger c_{j}^{\phantom{\dagger}}$ counts whether there is a fermion on site $j$.
Hopping with an amplitude $t$ occurs between pairs $\av{jk}$ of nearest neighbors. 
The dipole-dipole interaction $V^d_{jk} = c_d \left[1-3(\hat{r}_{jk}\cdot \dv)^2 \right]/(r_{jk}/a)^3$ depends on the vector $\mathbf{r}_{jk}$ connecting sites $j$ and $k$, $\hat{r}_{jk} = \mathbf{r}_{jk}/r_{jk}$,  $a$ is the lattice constant, and $c_d$ sets the strength of the DDI.
An external field orients the dipoles along the direction $\dv$, given by the spherical angles $\theta$, $\phi$, see Fig.~\ref{fig:lattice}. 

Many-body effects have a significant effect on the Lifshitz transition, resulting e.g.\ in its two-side character in three dimensions~\cite{Katsnelson2000} and interaction-driven band flattening in two dimensions~\cite{Irkhin02,Yudin14}, and therefore need to be properly taken into account. In order to achieve this goal, we employ the dual boson approach to strongly correlated systems~\cite{Rubtsov12,vanLoon14-2}, since it is capable of accounting for many-body effects in the strongly-interacting regime. This method has previously been applied to the dipolar Fermi-Hubbard model (DFH)~\cite{vanLoon15}, see Appendix~\ref{sec:appendixA} for additional computational details. 

In order to observe the Lifshitz transition, we start with a system below but close to half-filling, such that the Fermi surface is close to the VHSs and that even moderate deformations suffice to cross them. We use a density $n=0.40 \pm 0.01$. Our simulations use the grand canonical ensemble and therefore operate at fixed chemical potential. As a result, the particle density cannot be completely fixed. However, by subtracting the Hartree contributions to the chemical potential, the changes in density can be made negligible. 

Three relevant energy scales of this problem are given by the hopping amplitude, $t$, the temperature of the fermions, $T$, and the dipolar interaction strength, $c_d$. In order to observe a relatively sharp Fermi surface, the magnitude of $k_B T$, with $k_B$ Boltzmann's constant, has to be small compared to the hopping bandwidth of $8t$.  The dipolar coupling, in turn, determines the magnitude of the anisotropic effects. For experiments with highly magnetic lantanide  atoms, $t$ can be tuned over a wide range from hundreds of mHz to hundreds of Hz, while the dipolar coupling $c_d$ is set by the atomic species and the lattice spacing a selected. For Erbium with $a=266$ nm, $c_d$ was measured to be 40Hz~\cite{Baier201}.

A single component Fermi gas of highly magnetic atoms also offers an unprecedented and highly efficient cooling mechanism as DDI ensures a finite scattering cross-section and thus allows thermalization between atoms, while  the Pauli principle forbids short-range s-wave scattering and thus suppresses losses caused by inelastic three body collisions \cite{LuPRL12, AikawaPRL14}.
Efficient cooling is crucial when simulating condensed matter systems since the Fermi temperature $T_F$ changes from the Kelvin scale in solid state systems to the nano-Kelvin scale in the atomic gas.
In the bulk, temperatures down to $\approx10\%$ of the Fermi temperature have already been achieved for a polarized fermionic gas of highly magnetic atoms using the exceptional direct cooling possibility offered by DDI described above~\cite{AikawaPRL14}.
In the presence of a periodic potential, the Fermi temperature is set by half the bandwidth $T_F\approx 4t$.  
By minimizing heating effects, one can expect to keep $T/T_F$ nearly constant during the ramping up of the optical potential while in the Fermi Liquid regime \cite{Lode:2008,Dolfi:2015,Sheikhan:2015}.

Below, we exemplify the calculations by considering $t = 100$ Hz, $T = 20$ Hz, and $c_d = 50$ Hz, with all energies given in units of $t$. While this order of magnitude of $T/T_F$ has been achieved in experiments with ultracold Er in a harmonic trap~\cite{AikawaPRL14}, heating effects will need to be minimized in order to achieve a similar temperature in a lattice. In general, the lower the temperature, the sharper the Fermi surface and the clearer the Lifshitz transition can be observed.

\section{The Lifshitz transition}
\label{sec:Lifshitz}
\begin{figure}[t]
 \includegraphics[width=0.35\textwidth]{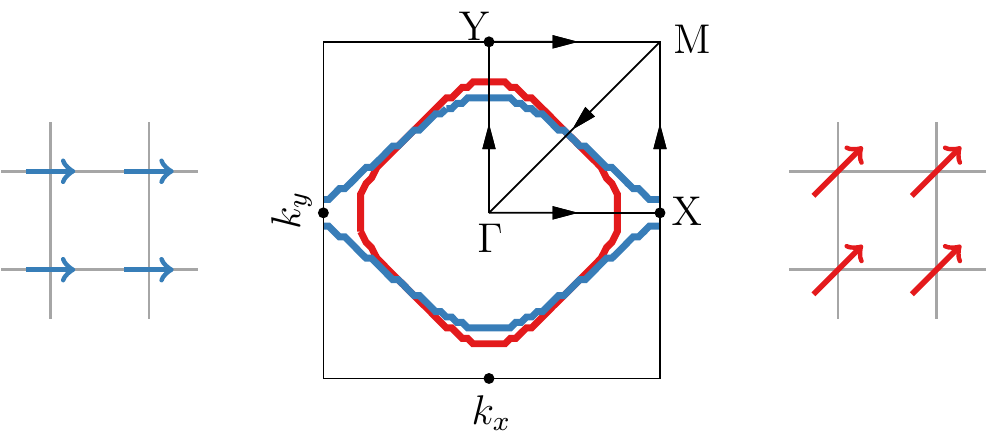}
 \caption{Fermi surfaces. The red curve corresponds to $\phi=0.25\pi$, where the Fermi surface is independent of the angle $\theta$. At $\phi=0$, $\theta=0.5\pi$ (blue line), the Fermi surface is deformed anisotropically due to the DDI. 
 This breaks the symmetry between the $X$ and $Y$ points in the Brillouin Zone.
 In the $k_x$ direction, the Fermi surfaces of neighboring Brillouin Zones are connected and the VHS (black dots) is enclosed by the Fermi surface. In the $k_y$ direction, the Fermi surfaces are not connected and the VHS is outside of the Fermi surface. 
 }
 \label{fig:FS}
\end{figure}

In a 2D square lattice, the Brillouin Zone (BZ) defines quasi-momenta $k_x, k_y \in [-\pi,\pi]$ in units of the inverse   lattice spacing. The stationary points of the dispersion $t_\kv = -2t \left[ \cos(k_x)+\cos(k_y)\right]$ are at the points $(\pm\pi,0)$ and $(0,\pm\pi)$ on the edge of the BZ (dots in Fig \ref{fig:FS}), corresponding to the VHSs of the non-interacting system.
The dipolar interaction does not affect the location of the VHSs.

In the absence of interactions, the $x$ and $y$ directions of the system are identical the Fermi surface resembles a diamond with rounded corners.  When the DDI is turned on, with dipoles oriented along the $xy$-diagonal ($\phi=0.25\pi$, any $\theta$), the Fermi surface preserves this shape, see Fig.~\ref{fig:FS} (red line). 

However, orienting the dipoles along the $x$-axis ($\phi=0$, $\theta=0.5\pi$), breaks the symmetry between the $x$ and $y$ directions. As a consequence, the Fermi surface  loses its symmetry as well. The resulting deformation leads to the Lifshitz transition: the Fermi surface now encloses the VHSs at $X=(q_x =\pm \pi,\,\,\, q_y = 0)$. Furthermore, the Lifshitz transition changes the topology of the Fermi surface, which now connects neighboring Brillouin Zones in the horizontal direction, as can be inferred from the periodic continuation of Fig.~\ref{fig:FS}.

An additional insight into the Lifshitz transition can be obtained by studying the properties of the spectral function, $A(E,k)$, which describes the energies and momenta of the single-particle excitations in the system. In order to highlight the anisotropy due to the DDI, we calculate the spectral function along two distinct paths in the Brillouin Zone, $\Gamma$-X-M and $\Gamma$-Y-M,  where $\Gamma=(0,0)$ is the origin and M$=(\pi,\pi)$ is the corner of the Brillouin Zone, as shown in  Fig.~\ref{fig:FS}. 

\begin{figure}[t]
 \includegraphics[width=0.48\textwidth]{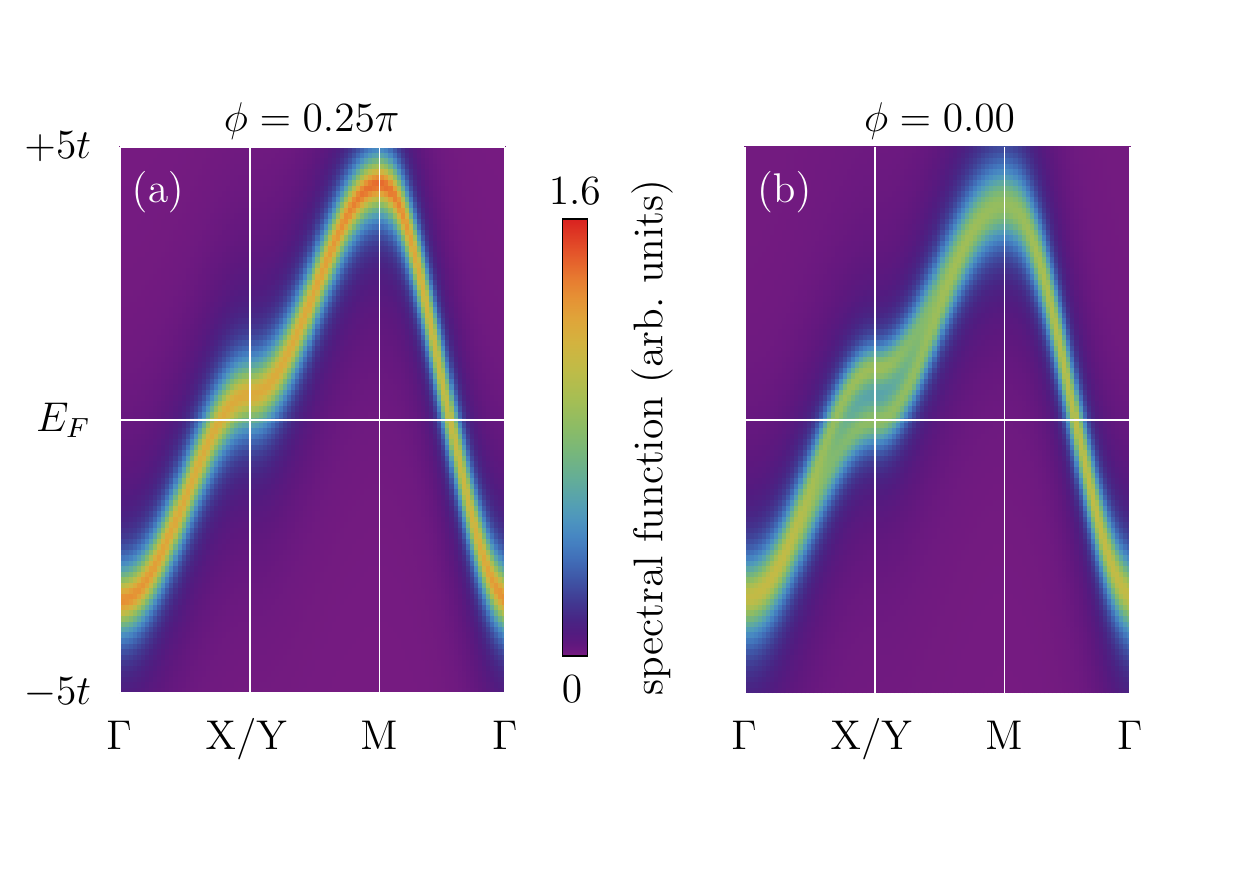}
 \caption{Spectral function at $\phi=0.25\pi$ (left) and $\phi=0$ (right), both for $\theta=0.50\pi$. The sum of the spectral function along the $\Gamma$-X-M and $\Gamma$-Y-M paths (see Fig.~\ref{fig:FS}) is shown. At $\phi=0.25\pi$, the spectral functions along these paths  are identical, whereas at $\phi=0$ the X and Y points are distinguishable due to the anisotropic interaction. This corresponds to the splitting into two bands. 
 }
 \label{fig:specfunc}
\end{figure}

Fig.~\ref{fig:specfunc} shows the sum of the two spectral functions along these paths for two different dipole orientations, with the Fermi energy at $E=0$ (white horizontal line). 
In the left panel of Fig.~\ref{fig:specfunc} we show the ``symmetric'' case $\phi=0.25\pi$ with $n=0.4<0.5$. In the absence of interactions, the Van Hove singularity crosses the Fermi surface exactly at half-filling.  In Fig.~\ref{fig:specfunc}(a), the VHS is clearly visible as a very flat dispersion at the $X$ and $Y$ points, however it is now located above the Fermi energy $E=0$.
Fig.~\ref{fig:specfunc}(b) shows the spectral function at $\phi=0$, on the other side of the Lifshitz transition.
Here, the dispersion has two branches corresponding to the X point and Y point respectively. The branches remain flat, corresponding to two VHSs, one above and one below the Fermi energy, see Appendix~\ref{sec:appendixB} for additional details.

A naive estimate of the energy difference between the $X$ and $Y$ points can be obtained using Hartree-Fock theory. 
As shown in  Appendix~\ref{sec:appendixB}, the contribution of the Fock diagram lowers (raises) the energy of the VHS at the X (Y) point by $\Delta E\approx 1.2 c_d$ in the limit of zero temperature and taking into account only  nearest-neighbor interaction.
This correctly predicts the order of magnitude of the splitting observed in Fig.~\ref{fig:specfunc}.
The interaction strength $c_d$ determines the scale of the anisotropy, therefore the energy-resolved measurements need a resolution of the same order to be able to detect these effects. In situations where the temperature is substantially larger than $c_d$, all effects are likely to be thermally smeared out. 

\begin{figure}[b]
 \includegraphics[width=0.48\textwidth]{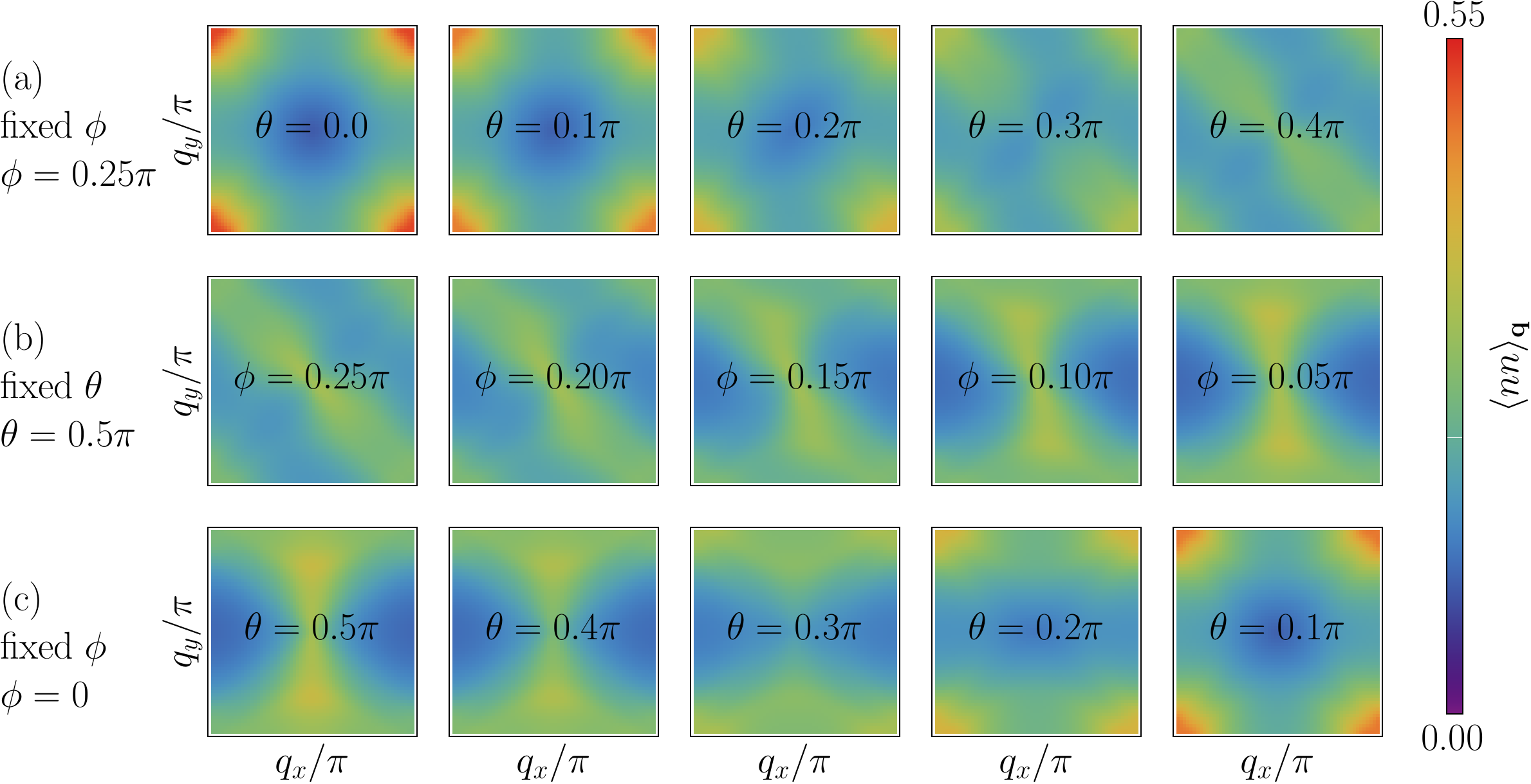}
 \caption{Static density-density correlation function in momentum space. The dipolar angles correspond to those of Fig.~\ref{fig:lattice}.
 }
 \label{fig:xlat}
\end{figure}

The static susceptibility $\av{nn}_\qv$, which is defined as the Fourier transform of the density-density correlation function to momentum space,  provides an alternative way to investigate the system.
Compared to the spectral function, which contains information on the single-particle excitations, the susceptibility gives access to the collective excitations, in particular, to the charge density waves. 
Thus, the susceptibility reveals whether the system is in a charge-ordered state.
Along with deforming the Fermi surface and altering the spectral function, the DDI  also affect the susceptibility, as demonstrated by Fig.~\ref{fig:xlat}.
For $\phi=0.25 \pi$ and dipoles perpendicular to the lattice plane, panel (a), we observe an isotropic susceptibility with maxima close to the $M=(\pm \pi,\pm \pi)$ points, which  corresponds to a checkerboard pattern in real space as the interaction is isotropically repulsive in plane.
As the dipoles get oriented parallel to the lattice plane while keeping $\phi=0.25\pi$, the symmetry between the two diagonals is broken, reflecting the  asymmetry between the direction $\phi=\pm 0.25\pi$ introduced by the anisotropy of DDI. For large $\theta$, a maximum starts to appear at small $q$ and long wavelength, which is reminiscent of the susceptibility observed in the ultralong-range ordered phase of the dipolar Fermi-Hubbard model~\cite{vanLoon15}.
Note that this evolution of the susceptibility is completely interaction-driven and happens while the Fermi surface remains unperturbed, as shown in Fig.~\ref{fig:FS}.
In panel (b), the effect of rotation in plane is illustrated, going from dipoles oriented along the diagonal ($\phi=0.25 \pi$) to dipoles pointing along the $x$-axis ($\phi=0$). We observe that the line of maxima in the susceptibility follows the dipole orientation angle, and the Fermi surface is deformed in this process, cf. Fig.~\ref{fig:FS}.
Finally, in Fig.~\ref{fig:xlat}c, we consider the path backwards to the dipoles aligned out of plane, now keeping $\phi=0$ constant. As for fixed $\phi=0.25\pi$, the susceptibility evolves from anisotropic to isotropic. However the orientation of the line of maxima is now rotated and in contrast to fixed $\phi=0.25\pi$, this evolution is associated to a deformation of the Fermi surface: at $\theta=0$ it is isotropic whereas at $\theta=0.5\pi$ it is deformed. 

\begin{figure}
 \includegraphics[width=0.48\textwidth]{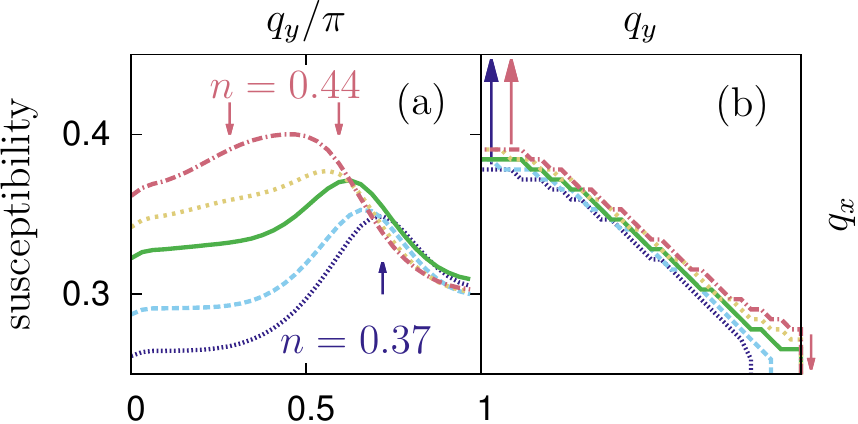}
 \caption{%
 (a) $q_x=0$ cross section of the static susceptibility. The arrows indicate the momenta corresponding to the transitions illustrated on the right.
 (b) Fermi surface (only the top right quadrant of the Brillouin Zone is shown), the green Fermi surface corresponds to Fig.~\ref{fig:FS}. Different lines correspond to the densities $n=$ $0.37$, $0.38$, $0.40$, $0.42$ and $0.44$. The dipole orientation is fixed to $\phi=0$, $\theta=0.5\pi$.
 }
 \label{fig:x_cross}
\end{figure}

Let us have a more detailed look onto the specific case of $\phi=0$, $\theta=0.5$, where the Fermi surface deformation is the largest. In Fig.~\ref{fig:x_cross}(a), we show a cross-section at $q_x=0$ of the susceptibility $\av{nn}_\qv$.
The green line corresponds to the same density, $n=0.40$, as in Fig.~\ref{fig:FS}.
The susceptibility changes, however, if one changes the  density. At the lowest density shown, $n=0.37$ (blue) there is a clear maximum in the susceptibility. 
This is the Kohn anomaly~\cite{Kohn59} corresponding to excitations from the flat top of the Fermi surface to the bottom of the next Fermi surface, as also shown in Fig.~\ref{fig:x_cross}(b) (blue arrow).

As the density increases, the Fermi surface expands (Fig. \ref{fig:x_cross}(b)) and the Kohn anomaly shifts to slightly lower momentum. 
When reaching a given critical density (here between $n=0.38$ and $0.40$), the Lifshitz transition occurs at the $X$-point in the Brillouin Zone. 
As a result, excitations with small momentum transfer are possible, as illustrated in Fig. \ref{fig:x_cross}(b) (arrows), and the susceptibility at small $q_y$ is greatly enhanced.
This time, instead of a sharp peak, there is a much broader enhancement. 
Since the $X$-point is a VHS, the single-particle energy close to $X$ only depends weakly on momentum, and so does the occupation $n_\kv$. This means that the density profile is relatively flat near the Fermi surface here and the corresponding excitations are less sharply peaked.

\section{Conclusions}

In this work we demonstrated that ultracold dipolar fermions in an optical lattice can be used as an efficient quantum simulation platform to study topological Lifshitz transitions. As their crucial property, the Lifshitz transitions predicted in ultracold quantum gases occur solely due to the anisotropic interparticle interactions, and therefore can be observed in an isotropic optical lattice.  It was shown that the transition can be detected by measuring the Fermi-surface deformations, the spectral function, and the static susceptibility. Thus, several complimentary experimental techniques can be used. 
The Fermi surface deformation can be determined using adiabatic mapping time-of-flight measurements~\cite{Greiner:2001,Koehl:2005}.  The spectral function can be revealed using momentum-resolved radiofrequency spectroscopy~\cite{Stewart08} or momentum-resolved Bragg scattering~\cite{Stenger:1999,Rey:2005,Ernst09}.
Lattice-modulation spectroscopy~\cite{Stoeferle:2004,Kollath:2006,Clark:2006} can show the energies of the available states, however without the momentum resolution. The splitting of the Van Hove singularity into two energies associated with the X and Y point can be investigated in this way.
The static susceptibility $\av{nn}_\qv$ can be accessed by two-body correlation analysis of the time-of-flight density distribution, so called noise-correlation measurement \cite{Altman:2004,Folling:2005,Rom:2006}.
These observation techniques will have to be integrated into the experimental set-up required for the Lifshitz transition.

In order to observe the interaction-induced Lifshitz transition, the fermion density needs to be close to the Van Hove filling, which for nearest-neighbor hopping occurs at half-filling. Furthermore, since the phase transition point  depends on the local density, the confinement potential has to be be sufficiently flat to simultaneously induce the Lifshitz transition in a large part of the trap. 
Furthermore, Fermi surface deformations are most naturally observed in momentum space, so observation is helped by homogeneity. 
Novel techniques such as box traps~\cite{Gaunt13} and anticonfinement potentials~\cite{Ma08,Mathy12} may help in reducing inhomogeneous trapping effects. Other techniques such as single-site adressing \cite{Bakr10,Sherson10,Weitenberg11,Cheuk15,Parsons15,Haller15,Edge15,Omran15} or super-lattice  engineering/tuning \cite{Sebby-Strabley06,Folling07} may help both in preparing regions of controlled filling and give access to original probing schemes \cite{Trotzky10,Endres2013}.

While ultracold magnetic atoms are the primary candidates to observe the interaction-induced Lifshitz transitions, similar measurements can be performed with high-density samples of ultracold heteronuclear molecules~\cite{OspelkausFDiss09, NiNature10, ParkPRL15, Park15b}.
Furthermore, the phenomenon is expected to occur for other types of anisotropic interparticle interactions, such as quadrupole-quadrupole couplings~\cite{BhongalePRL13, LarzPRA13} or interactions induced by far-off-resonant laser fields~\cite{LemeshkoPRA11Optical, LemFri11OpticalLong}.

\acknowledgments

We thank Francesca Ferlaino, Koen Reijnders and Jan Kaczmarczyk for useful discussions. E.G.C.P.v.L. and M.I.K. acknowledge support from ERC Advanced Grant 338957 FEMTO/NANO. L.C. acknowledges support by the FWF through SFB FoQuS and START grant under Project No. Y479-N20.

\appendix

\section{The dual boson approach}
\label{sec:appendixA}

We use the dual boson formalism to strongly correlated systems~\cite{Rubtsov12,vanLoon14-2}. Here, we give a short synopsis of the method. The main idea of the approach is to separate the interaction effects into two parts: momentum-independent mean-field effects and momentum-dependent corrections. 

The first stage of the computation is the determination of the effective mean-fields. This is achieved by introducing an auxiliary single-site problem with \emph{dynamical}, local fields $\Delta_\nu$, $\Lambda_\omega$, that replace the non-local terms $t_{jk}$ and $V_{jk}$ of the original system. In the action formulation, this auxiliary problem is defined as
\begin{align}
 S=- \sum_{\nu} c^\ast_{\nu} \left[i\nu+\mu-\Delta_\nu \right] c^{\phantom{\ast}}_{\nu} + \frac{1}{2} \sum_\omega \Lambda_\omega n_\omega n_\omega,
\end{align}
where $\nu$ and $\omega$ are fermionic and bosonic Matsubara frequencies respectively.
This single-site problem can be solved numerically exactly, and the Green's function and two-particle correlation functions can be determined. Our numerical solution of the auxiliary single-site problem is based on the ALPS libraries~\cite{ALPS2,Hafermann13,Hafermann14}.
From the Green's function and susceptibility of the auxiliary model, we then obtain an approximation for the Green's function and susceptibility of the lattice model.
The fields $\Delta_\nu$ and $\Lambda_\omega$ are chosen self-consistently, by requiring the local Green's function and susceptibility of the lattice model to be identical to those of the auxiliary problem.

\begin{figure}[h]
 (a)
\begin{tikzpicture}
  \coordinate (diagram1) at (0,0) ;

\coordinate (leftend) at ($(diagram1)$ ) ;
\coordinate (leftfermion) at ($(leftend) + (-0.3,0)$) ;
\coordinate (toplefttriangle) at ($(leftend) + (0.3,0.4)$) ;
\coordinate (rightlefttriangle) at ($(leftend) + (0.6,0.)$) ;

\coordinate (leftrighttriangle) at ($(rightlefttriangle) + (0.7,0)$) ;
\coordinate (toprighttriangle) at ($(leftrighttriangle) + (0.3,0.4)$) ;
\coordinate (rightend) at ($(leftrighttriangle) + (0.6,0.)$) ;
\coordinate (rightfermion) at ($(rightend) + (0.3,0)$) ;

\draw[very thick,fill=red!50] (leftend) -- (toplefttriangle) -- (rightlefttriangle) -- cycle;

\draw[very thick,fill=red!50] (rightend) -- (toprighttriangle) -- (leftrighttriangle) -- cycle;

\draw[thick,->-=0.5] (leftfermion) -- (leftend) ;
\draw[thick,-<-=0.5] (rightfermion) -- (rightend) ;

\draw[thick,->-=0.5] (rightlefttriangle) -- (leftrighttriangle);
\draw[thick,decorate,decoration=snake] (toprighttriangle) -- (toplefttriangle);
\end{tikzpicture}
\,\,\,(b)
\begin{tikzpicture}
    \coordinate (diagram1) at (0,0) ;

    \coordinate (leftend) at ($(diagram1)$ ) ;
    \coordinate (leftboson) at ($(leftend) + (-0.3,0)$) ;
    \coordinate (toplefttriangle) at ($(leftend) + (0.4,0.3)$) ;
    \coordinate (bottomlefttriangle) at ($(leftend) + (0.4,-0.3)$) ;

    \coordinate (toprighttriangle) at ($(toplefttriangle) + (.7,0)$) ;
    \coordinate (bottomrighttriangle) at ($(toprighttriangle) + (0,-0.6)$) ;
    \coordinate (rightend) at ($(toprighttriangle) + (0.4,-0.3)$) ;
    \coordinate (rightboson) at ($(rightend) + (0.3,0)$) ;

    \draw[very thick] (leftend) -- (toplefttriangle) -- (bottomlefttriangle) -- cycle;

    \draw[very thick,fill=red!50] (rightend) -- (toprighttriangle) -- (bottomrighttriangle) -- cycle;

    \draw[thick,decorate,decoration=snake] (leftboson) -- (leftend) ;
    \draw[thick,decorate,decoration=snake] (rightboson) -- (rightend) ;

    \draw[thick,-<-=0.5] (bottomlefttriangle) -- (bottomrighttriangle);
    \draw[thick,-<-=0.5] (toprighttriangle) -- (toplefttriangle);
\end{tikzpicture}
\caption{Feynman diagrams employed in the dual boson approach. Diagram (a) renormalizes the fermion propagator and diagram (b) renormalizes the susceptibility. The Fermi surface deformation occurs due to the anisotropy of diagram (a), coming from the anisotropic DDI.}
\label{fig:feynmandiagrams}
\end{figure}
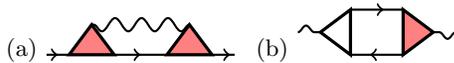

The second stage consists of momentum-dependent ``dual'' corrections to the mean-field solution. These are crucial for studying Fermi surface deformations, since that is an essentially momentum-dependent phenomenon. The associated diagrams are shown in Fig.~\ref{fig:feynmandiagrams}, we refer the reader to Ref.~\onlinecite{vanLoon14-2} for explicit formulas. In these diagrams, the lines with arrows describe fermion propagation, the wiggly lines the propagation of density fluctuations and the (filled) triangles the (ladder-renormalized~\cite{vanLoon14-2}) interaction between the fermions and the density fluctuations. The numerical values of these elements are determined from the auxiliary model.

The Fock-like diagram in Fig.~\ref{fig:feynmandiagrams}(a) is essential to the Fermi surface deformation. Due to the DDI, the wiggly line is anisotropic and as a result, the self-energy is also anisotropic and the Fermi surface deforms. The dual diagrammatic technique was applied until (inner) self-consistency~\cite{vanLoon14-2} was achieved (usually 10 iterations were sufficient), to allow for feedback of the Fermi surface deformation on the susceptibility via diagram~\ref{fig:feynmandiagrams}(b) and back. Finally, the nonlocal corrections from the dual diagrammatic technique are applied to the original fermions.

The calculations were performed on a $64 \times 64$ square lattice. The Fermi surface is determined from the Green's function at the point where the occupation $n_\kv$ crosses $1/2$. There is a small discretization uncertainty due to the finite momentum resolution. The spectral function of Fig.~\ref{fig:specfunc} was obtained from the Green's function at Matsubara frequencies using Pad\'e approximants~\cite{Vidberg77}.

\section{Estimates at zero temperature}
\label{sec:appendixB}

In order to get a feeling for the expected magnitude of the effects, here we perform a perturbative analysis. 
It is most convenient to do this at $T=0$ and close to half-filling, where the integrals over the Brillouin Zone can be drastically simplified.

\begin{figure}[t]
 \includegraphics[width=0.2\textwidth]{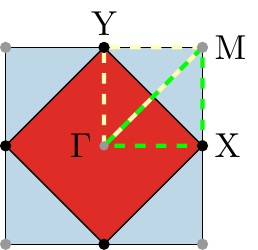}
  \caption{For a non-interacting, half-filled system, the red part of the Brillouin Zone lies within the Fermi surface. The black and gray dots denote the Van Hove singularities (stationary points and end points of the dispersion respectively), and we define the points $\Gamma=(0,0)$, X$=(\pi,0)$, Y$=(0,\pi)$ and $M=(\pi,\pi)$. }
  \label{fig:bz}
 \end{figure}

The energy of the non-interacting system is given by the Fourier transform, $t_\kv$, of the hopping,
\begin{align}
 H_0 =&~ \sum_\kv E_\kv n_\kv \notag \\
 E_\kv =& ~t_\kv = -2t \left[ \cos(k_x)+\cos(k_y)\right] \notag
\end{align}
The Fermi surface is determined by the condition of  $E_\kv=\mu$, with $\mu$ the chemical potential. Half-filling occurs at $\mu=0$, and the resulting Fermi surface the diamond shown in Fig.~\ref{fig:bz}. In the main text, we studied a system below half-filling at $n\approx 0.4$, where the Fermi surface is slightly smaller than the diamond. 

The Van Hove singularities (black dots) are found as the stationary points of the dispersion, $\nabla_\kv E_\kv = 0$. Two saddle points occur at the center of the sides of the Brillouin Zone. The global minimum and maxima of the dispersion are shown as the gray dots, at the origin and the corners of the Brillouin Zone respectively. Accounting for the periodicity, there are two saddle points, one minimum and one maximum per Brillouin Zone, the minimum number of critical points predicted by Van Hove~\cite{VanHove53}.

Let us now  consider an anisotropic interaction. For simplicity, we take into account only the nearest-neighbor couplings and set the dipoles' orientation along the $x$-axis ($\theta=\pi/2$, $\phi=0$). In momentum space, that interaction is given by
\begin{align}
 V_\qv =& \,\, 2c_d \left[-2 \cos(q_x) + \cos(q_y)\right].  
\end{align}
Now, we estimate the self-energy of the fermion using the Hartree-Fock approximation.~\cite{Mahan00}
The expectation value with respect to $H_0$ is denoted by $\av{\cdot}^0$.
The Hartree contribution to the self-energy is independent of $\kv$ and only leads to a change in the chemical potential, which can be ignored.
The Fock contribution, on the other hand, induces anisotropy~\cite{Miyakawa08}:
\begin{align}
\Sigma^{\text{Fock}}_\kv =& -\sum_\qv V_\qv \av{n_{\kv+\qv}}^{0} \label{eq:newenergy}.
\end{align}
The sums over momenta in equation~\eqref{eq:newenergy} should be understood as normalized integrals over the Brillouin Zone.

Performing this calculation explicitly for the high-symmetry points $Y$ and $X$, i.e., $\kv=(0,\pi)$ and $\kv=(\pi,0)$, reveals the anisotropy. The energy of the $Y$ point with respect to the Fermi energy is given by:
\begin{align}
 \Sigma^{\text{Fock}}_{\kv=Y} =& -\frac{1}{(2\pi)^2} \int\limits_{\kv+\qv \in \text{Fermi volume}} 
 \!\!\!\!\!  \!\!\!\!\!  \!\!\!\!\!  \!\!   
 V_\qv \, d\qv \notag \\
 =& 12~c_d /\pi^2 \notag \\
 \approx& 1.2~ c_d
\end{align}
On the other hand, for the $X$ point, $\Sigma^{\text{Fock}}_{\kv=X}\approx -1.2~ c_d$, and the dispersion is pushed below the Fermi energy. Here we used that $\av{n}^{0}$ is zero outside of the Fermi surface and unity inside. These estimates of the energy splitting between the $X$ and $Y$ points match the order of magnitude of the results in Fig.~\ref{fig:specfunc}. Note that an exact match is not expected, since the results of Fig.~\ref{fig:specfunc} are obtained at finite temperature, away from half-filling and with interaction beyond nearest neighbors.

In Sec.~\ref{sec:Lifshitz}, the numerical results showed that the VHSs do not move in the presence of interaction. 
This can be demonstrated perturbatively in the zero-temperature limit.
Let us show that the energy \eqref{eq:newenergy} is stationary at these points. The first term, $t_\kv$, is stationary since these points are the VHSs of the non-interacting system.
Then, we have to determine the gradient of $\av{n_{\kv+\qv}}^{0}$. 
Since the density is a step function, its derivative is a delta function on the Fermi surface.
\begin{align}
 \nabla_\kv \left(\Sigma^{\text{Fock}}_{\kv}\right) =& -\nabla_\kv \sum_\qv V_\qv \av{n_{\kv+\qv}}^{0} \notag \\
 =& \int\limits_{\kv+\qv\in \text{Fermi surface}} 
 \!\!\!\!\!  \!\!\!\!\!  \!\!\!\!\!  \!\!   
 -V_\qv \, d\qv \\
 \nabla_\kv \left(\Sigma^{\text{Fock}}_{\kv}\right)|_{\text{X, Y}}=& \,\,\, 0 \notag
\end{align}
Since the gradient of the Fock self-energy is zero at the $X, Y$ points, they are also the stationary points of the interacting system.

\bibliography{lifshitz}

\end{document}